\documentclass[12pt]{article}

\usepackage{parskip}  
\usepackage{mathtools}
\usepackage{amsfonts}
\usepackage{amssymb}
\usepackage{amsthm}
\usepackage{amsmath}
\usepackage[utf8]{inputenc} 
\usepackage[svgnames]{xcolor}
\usepackage[colorlinks]{hyperref} 
\usepackage{tikz}       
\usepackage{fancyhdr}   
\usepackage{titlesec}   
\usepackage{times}
\usepackage{changepage}
\usepackage{graphicx}
\usepackage{float}
\usepackage{bbm}        
\usepackage{pdfpages}
\usepackage{enumitem}   
\usepackage{caption}    
\usepackage{subcaption} 
\usepackage{courier}    
\usepackage{adjustbox}  
\usepackage{booktabs}
\usepackage{multirow}
\usepackage{colortbl}   
\usepackage{authblk}
\usepackage[round]{natbib}
\usepackage[nottoc,numbib]{tocbibind}  

\titlelabel{\thetitle.\;}             
\newtheorem{assumption}{Assumption}

\hypersetup{
	citecolor  = {gray},
	colorlinks = true,
	linkcolor  = blue
}        

\hypersetup{pdfborder={0 0 0}}  
\theoremstyle{plain}

\newcommand*\dif{\mathop{}\!\mathrm{d}}

\title{ \vspace{-1.5\baselineskip}
	\bf A Comparative Review of Specification Tests for Diffusion Models
}
\author[1]{L\'opez-P\'erez, A. \thanks{\textit{Contact:} \href{mailto:alejandralopez.perez@usc.es}{alejandralopez.perez@usc.es}. The authors gratefully thank Spanish National Research Council for the computing resources of the Supercomputing Center of Galicia (CESGA). The authors acknowledge support from grant PID2020-116587GB-I00 from the Spanish Ministry of Economy and Competitiveness.} }
\author[1]{Febrero-Bande, M.}
\author[1]{Gonz\'alez-Manteiga, W.}
\affil[1]{Department of Statistics, Mathematical Analysis and Optimization. Universidade de Santiago de Compostela, Spain.}
\date{}

\begin{document}
	
\maketitle

\begin{center}
	\begin{minipage}{0.9\textwidth}
		\vskip 0.05cm
		\small 
		\textbf{Abstract.} Diffusion models play an essential role in modeling continuous-time stochastic processes in the financial field. Therefore, several proposals have been developed in the last decades to test the specification of stochastic differential equations. We provide a survey to collect some developments on goodness-of-fit tests for diffusion models and implement these methods to illustrate their finite sample behavior, regarding size and power, by means of a simulation study. We also apply the ideas of distance correlation for testing independence to propose a test for the parametric specification of diffusion models, comparing its performance with the other methods and analyzing the effect of the curse of dimensionality. As real data examples, treasury securities with different maturities are considered.



%
%

\vskip 1em
\textbf{Keywords.} Diffusion process; Distance correlation; Goodness-of-fit; Stochastic differential equations.

\textbf{JEL codes.} C12; C22; C52.
\end{minipage}
\end{center}






\newpage

\section{Introduction}

Financial theory has been developing on a continuous-time basis, where the evolution of assets are usually represented in terms of continuous-time stochastic processes, rooted in the foundational works of \cite{black1973pricing} and \cite{merton1975asymptotic}. 
%
With market prices evolving at very short-time scales, continuous-time models have proved to be more appropriate to capture the quick changes observed on real markets than discrete-time models.
In this continuous-time paradigm, modeling interest rates represents one of the main topics in the recent financial economics literature, as it lies at the core of the most basic valuations problems in finance. The uncertainty attached to interest rates evolution is a relevant part of the theory of financial decision making, so that understanding the factors that drive interest rates is important for modeling the yield curve, analyzing financial instruments (futures, options on zero coupon bonds, swaps contracts, etc.) or developing hedging interest rate risk strategies.
One of the most common representation is the time-homogeneous diffusion given by a stochastic differential equation (SDE) driven by a Wiener process $\lbrace W_t \rbrace_{t \geq 0}$,
\begin{equation} \label{eq:SDEnoparam}
\dif X_t = m(X_t)\dif t + \sigma(X_t) \dif W_t,
\end{equation}
defined on a complete probability space $\big(\Omega, \mathcal{F}, \lbrace \mathcal{F}_t \rbrace_{t \in [0,T]}, \mathbb{P}\big)$, where $\Omega$ is a nonempty set, $\mathcal{F}$ is a right-continuous $\sigma$-algebra of subsets of $\Omega$ and $\mathbb{P}$ is a probability measure, $\mathbb{P}(\Omega) = 1$. The process $X_t \in \mathbb{R}$ evolves over the interval $[0,T]$ in continuous time, according to the drift $m(\cdot)$ and diffusion or volatility $\sigma(\cdot)$ functions.

The correct specification of the dynamics of short rates is fundamental to much of empirical and theoretical finance, as mis-specification could lead to pricing errors, mishedged or unhedged risks, or incorrect inferences. Over the last four decades, several developments of short-term interest rate models have emerged. \cite{chan1992empirical} developed a general framework to nest different single-factor models of the short-term interest rate, including those by \cite{merton1973theory}, \cite{cox1975notes}, \cite{cox1976valuation}, \cite{vasicek1977equilibrium}, \cite{Dothan1978}, \cite{brennan1980analyzing}, \cite{rendleman1980pricing}, \citeauthor{cox1980analysis} (\citeyear{cox1980analysis}, \citeyear{cox1985intertemporal}), or \cite{constantinides1982optimal}.
Given the profusion of diffusion-based models, the problem of choosing a family of models to explain the dynamics of interest rates arises naturally. 
Yet there is no consensus on the choice of the parametric specification of the short rate process.
With model diagnosis playing an essential role to determine which model best captures the short rates dynamics, there exist several proposals for continuous-time model specification with works dealing with goodness-of-fit test in different directions. 
\cite{Ait-Sahalia1996} proposed one of the first nonparametric test for univariate diffusion processes by comparing the model-implied parametric stationary density with a kernel density estimator based on discretely sampled observations, as the drift and diffusion functions completely characterize the stationary and marginal density. Years later, \cite{gao2004adaptive} developed a procedure to improve its finite sample performance. The comparison between parametric and nonparametric estimates of defining attributes of diffusion models has motivated different proposals, focusing on the marginal density  \citep{gao2004adaptive} and transitional density function (\citealp{hong2004nonparametric}; \citealp{chen2008test}), while \cite{corradi2005bootstrap} compared cumulative distribution functions.

A body of literature is focused in testing the drift or volatility functions of the diffusion processes, as an alternative to test the characterization of the whole dynamics of the process. 
With respect to the drift function, \cite{lee2006bickel} adapted the \cite{bickel1973some} test for diffusion models, \cite{arapis2006empirical} proposal used nonparametric estimation procedures, \citeauthor{negri2009goodness} (\citeyear{negri2009goodness}, \citeyear{negri2010goodness}) test is based on the score marked empirical process, \cite{kutoyants2010goodness} studied direct analogues of the classical Anderson-Darling and Kolmogorov-Smirnov tests and \cite{masuda2011goodness} used an innovation martingale approach.
Regarding the diffusion function, 
\cite{corradi1999specification} proposed a homoscedasticity test; 
\cite{dette2003test}, \cite{dette2008testing} and \cite{podolskij2008range} tests are based on stochastic processes of the integrated volatility; 
\cite{li2007testing} and 
\cite{dette2006estimation} rely on the $L_2$-distance between the diffusion function under the null and the alternative hypotheses;
\cite{zheng2009testing} and \cite{chen2019nonparametric} developed nonparametric tests.

Lastly, the specification of both drift and volatility functions is analyzed in
\cite{fan2003reexamination}, \cite{fan2003time} and \cite{ait2009nonparametric}, based on generalized likelihood ratio test ideas \citep{fan2001generalized};
\cite{gao2008specification} considered smoothing techniques for semiparametric diffusion models;
\cite{chen2010characteristic} used conditional characteristic function characterization to construct a specification test, while
\cite{kristensen2011semi} relied on the transition density; 
\cite{song2011martingale} test is based on the infinitesimal operator;
\cite{chen2011simultaneous} used an empirical likelihood statistic; and 
\cite{monsalve2011goodness} and \cite{chen2015asymptotically} tests are based on empirical regression processes.




In the present paper, we provide a survey to collect these developments in goodness-of-fit tests for diffusion models. Several of these proposals are extensions of earlier works to the context of continuous-time series, and consequently their drawbacks are also inherited. Due to the curse of dimensionality most of the existing tests cannot be easily extended to multivariate models. For tests based on nonparametric methods, bandwidth selection is a concern and power vitally depends on the dimension of the conditioning variables, decreasing with dimension. Though different adaptations have been addressed in order to avoid the curse of dimensionality, few have been studied for diffusion processes. As a sideways contribution, we apply distance correlation \citep{szekely2007measuring} ideas to test the specification of diffusion models, which circumvents the curse of dimensionality.


In a seminal series of works, \cite{szekely2007measuring} and \citeauthor{szekely2009brownian} (\citeyear{szekely2009brownian}, \citeyear{szekely2012uniqueness}, \citeyear{szekely2013distance}, \citeyear{szekely2013distance}) introduced the concepts of distance covariance and distance correlation as measures of the degree of dependence among vectors of random variables and used it to test independence. Since then, several works considered distance correlation to construct test statistics. \cite{zhou2012measuring} extended the concept to auto-distance correlation in time series to explore nonlinear dependence. \cite{davis2018applications} applied distance correlation ideas to stationary univariate and multivariate time series in order to measure lagged auto- and cross-dependencies. \cite{matsui2017distance} extended distance covariance notions to continuous-time stochastic processes defined on the same interval, while \cite{dehling2020distance} built a measure of independence for the components of i.i.d. pairs of stochastic processes on the unit interval and established consistency of the bootstrap procedure for the sample distance covariance/correlation of two processes. \cite{kroll2021asymptotic} derived the asymptotic distribution of sample distance covariance under mixing conditions.



The paper is organized as follows. Section~\ref{sec:2} provides a survey of goodness-of-fit test proposals for diffusion models and Section~\ref{sec:3} concerns the extension of distance correlation ideas to test the specification of diffusion processes. The implementation of some of the methods and the study of the level and power of the tests for finite samples are discussed in Section~\ref{sec:4} by means of Monte Carlo simulations and real data are analyzed in Section~\ref{sec:5}. Concluding remarks are provided in Section~\ref{sec:6}.

\section{Goodness-of-fit test for diffusion models} \label{sec:2}

In this section, we discuss the goodness-of-fit testing problem of diffusion models and introduce different methodologies developed in order to test the specification of the model, many of which are extensions of classical tests ideas from earlier works. We begin by introducing the parametric version of the SDE in~\eqref{eq:SDEnoparam},
\begin{equation} \label{eq:SDEparam}
\dif X_t = m(X_t, \boldsymbol{\theta})\dif t + \sigma(X_t, \boldsymbol{\theta}) \dif W_t.
\end{equation}
Under a parametric framework, $\boldsymbol{\theta}$ is an unknown parameter vector such that $\boldsymbol{\theta} \in \Theta \subset \mathbb{R}^d$ with $d$ a positive integer and $\Theta$ a compact parametric space, and ${m(\cdot, \boldsymbol{\theta})\colon \mathbb{R} \times \Theta \rightarrow \mathbb{R}}$ and ${\sigma(\cdot, \boldsymbol{\theta}) \colon \mathbb{R} \times \Theta \rightarrow (0,\infty)}$. A unique strong solution $\lbrace X_t \rbrace_{t \in [0,T]}$ of the stochastic differential equation (SDE), with initial condition $X_0 = x_0 \in \mathbb{R}$ adapted to the filtration $\lbrace \mathcal{F}_t \rbrace_{t \in [0,T]}$,
\begin{equation*} \label{eq:SDEint}
X_t = X_{0} + \int_0^T m(X_u, \boldsymbol{\theta})\dif u + \int_0^T \sigma(X_u, \boldsymbol{\theta}) \dif W_u,
\end{equation*}
exists under the assumptions that both drift $m(\cdot)$ and volatility $\sigma(\cdot)$ functions satisfy global Lipschitz continuous and growth conditions (see, e.g., \citealp{karatzas2012brownian}).

\begin{assumption}[Global Lipschitz]
	For all $x, y \in \mathbb{R}$ there exist a constant $C_1 < \infty$ independent of $\boldsymbol{\theta}$ such that \[ \left|m(x,\boldsymbol{\theta})-m(y,\boldsymbol{\theta})\right| + \left|\sigma(x,\boldsymbol{\theta})-\sigma(y,\boldsymbol{\theta})\right| \leq C_1 \left|x-y\right|. \]
\end{assumption}
\begin{assumption}[Linear growth]
	For all $x, y \in \mathbb{R}$ there exist a constant $C_2 < \infty$ independent of $\boldsymbol{\theta}$ such that \[ \left|m(x,\boldsymbol{\theta})\right| + \left|\sigma(x,\boldsymbol{\theta})\right| \leq C_2 \left(1 + \left|x\right| \right). \]
\end{assumption}

Although the diffusion model is formulated in continuous time, data are registered in discrete time points. In order to obtain a discrete version of it, one of the most used approximation schemes is the Euler-\cite{maruyama1955continuous} method: given an Itô process $\left\lbrace X_t \right\rbrace_{0 \leq t \leq T} $, solution of the SDE in~\eqref{eq:SDEparam} with initial value $X_{t_0} = x_0$ and the discretization of the time interval $[0, T]$, at $n$ equally spaced discrete time points ${0=t_0 < t_1 < \dots < t_n = T}$, the Euler-Maruyama approximation of $X_t$ is a continuous stochastic process that satisfies the iterative scheme
\begin{equation} \label{eq:SDEdisc}
X_{t_{i+1}} - X_{t_{i}} \approx m(X_{t_i}, \boldsymbol{\theta}) (t_{i+1} - t_{i}) + \sigma(X_{t_i}, \boldsymbol{\theta}) (W_{t_{i+1}} - W_{t_i}),
\end{equation}
with $i=0,1,\dots,n-1$, $X_{t_0} = x_0 \in \mathbb{R}$ and $t_i = i\Delta$, where $\Delta$ is the length of the sampling interval, that is $\Delta = T/n = (t_{i+1} - t_{i})$.


\subsection{Test based on the estimation of the distribution and density functions}


Many classical goodness-of-fit (GoF) tests proposals are based on the comparison of functions that characterize the probability law given in the null hypothesis and an estimator of this function, where the characterization can take place with the distribution ($F$), density ($f$), quantile ($F^{-1}$) or characteristic function ($\varphi$). The GoF methodology is rooted in the comparison of a nonparametric pilot estimator of the distribution or density function of the i.i.d. observations of a random variable $X$ with a consistent estimator of the target function. That is, to test the null hypothesis for the distribution function,
\begin{equation} \label{eq:hypDistr}
\begin{cases}
\mathcal{H}_0 \colon F \in \mathcal{F}_{\text{dist}} = \lbrace F_{\boldsymbol{\theta}}, \boldsymbol{\theta} \in \Theta \rbrace, \\
\mathcal{H}_\text{a} \colon F \notin \mathcal{F}_{\text{dist}},
\end{cases}
\end{equation}
or the density function
\begin{equation*} 
\begin{cases}
\mathcal{H}_0 \colon f \in \mathcal{F}_{\text{den}} = \lbrace f_{\boldsymbol{\theta}}, \boldsymbol{\theta} \in \Theta \rbrace, \\
\mathcal{H}_\text{a} \colon f \notin \mathcal{F}_{\text{den}},
\end{cases}
\end{equation*}
where the pilot estimator is given by the empirical cumulative distribution function ($F_n$) and the kernel density estimator ($f_{nh}$), respectively. The construction of the test statistic $T_n$ is based on some distance between the pilot estimator ($F_n$, $f_{nh}$) and the estimator under the null hypothesis ($F_{\hat{\boldsymbol{\theta}}}$, $f_{\hat{\boldsymbol{\theta}}}$) of the true function ($F_0$, $f_0$). For the distribution function, we have the test statistic $T_n = D(F_n, F_{\hat{\boldsymbol{\theta}}})$ and the diversity of tests comes from the diversity of distances $D$. If we take
\begin{equation*}
T_n = n \int_{-\infty}^{\infty} H(x) \left[ F_n(x) - F_{\hat{\boldsymbol{\theta}}}(x) \right]^2 \dif F_{\hat{\boldsymbol{\theta}}} (x),
\end{equation*}
we have the Cramér-von Mises family of statistics. Taking $H(x) \equiv 1$ we obtain the Cramér-von Mises test, while the Anderson-Darling test is obtained with the weight function $H(x) = \left[F_{\hat{\boldsymbol{\theta}}}(x) \left(1 - F_{\hat{\boldsymbol{\theta}}} \right)\right]^{-1}$. Considering an uniform metric,
\begin{equation*}
T_n = \sup\limits_x \sqrt{n} \left| F_n (x) - F_{\hat{\boldsymbol{\theta}}}(x) \right|,
\end{equation*}
we have the Kolmogorov-Smirnov test.

Direct analogues of these test can be found for diffusion model, as the drift $m(\cdot)$ and diffusion $\sigma^2(\cdot)$ terms completely characterize the stationary (conditional) or marginal (unconditional) density of the process. To test the specification of the drift and/or volatility function, given a specific parametrization, we have a join parametric family
\begin{equation*}
\mathcal{P}_{\boldsymbol{\theta}} \equiv \big\lbrace 
\left( m(\cdot, \boldsymbol{\theta}), \sigma^2 (\cdot, \boldsymbol{\theta}) \right) 
\big/ \, \boldsymbol{\theta} \in \Theta \big\rbrace,
\end{equation*}
with $\Theta$ a compact subset of $\mathbb{R}^d$. The test addresses if the values of the parameters in $\Theta$ are a representation of the true process, that is, whether the true functions $\mu_0(\cdot)$ and $\sigma_0^2(\cdot)$ belong to the parametric space $\mathcal{P}_{\boldsymbol{\theta}}$. Regarding the unique characterization of the marginal and transitional densities, any parametrization of $\mathcal{P}_{\boldsymbol{\theta}}$ of $m(\cdot)$ and $\sigma^2(\cdot)$ corresponds to a parametrization of the density, 
\begin{equation*}
\Pi_{\boldsymbol{\theta}} \equiv \big\lbrace 
\left( \pi(\cdot, \boldsymbol{\theta}), p(\cdot, \cdot, \mid \cdot, \cdot; \boldsymbol{\theta}) \right)
\big/ \left( m(\cdot, \boldsymbol{\theta}), \sigma^2 (\cdot, \boldsymbol{\theta}) \right) \in \mathcal{P}_{\boldsymbol{\theta}},\boldsymbol{\theta} \in \Theta \big\rbrace,
\end{equation*}
where $\pi(x, \boldsymbol{\theta})$ is the marginal density at $x$ and $p(s, y \mid t, x; \boldsymbol{\theta})$ is the transitional density from $x$ at time $t$ to $y$ at time $s$. Therefore, to test the model specification, the mapping between the drift and volatility functions and the transitional and marginal densities can be used. The marginal density corresponding to the drift and diffusion functions pair $\big(m(\cdot, \boldsymbol{\theta}), \sigma^2(\cdot, \boldsymbol{\theta})\big)$ has a close form expression via Kolmogorov forward equation and is given by
\begin{equation*}
\pi (x, \boldsymbol{\theta}) = \frac{\xi(\boldsymbol{\theta})}{\sigma^2(x,\boldsymbol{\theta})} \exp \left( \int_{x_0}^x \frac{2 m(u,\boldsymbol{\theta})}{\sigma^2(u,\boldsymbol{\theta})} \dif u \right),
\end{equation*}
where $\xi(\boldsymbol{\theta})$ is a normalization constant, determined to insure that the density function integrates to one. However, the transition density as defined by the Kolmogorov backward equation may not admit a closed form expression (see \citeauthor{ait1999transition}, \citeyear{ait1999transition}, \citeyear{ait2002maximum}, for Edgeworth type approximations to overcome this problem). The kernel estimator for the marginal density is
\begin{equation} \label{eq:pi_nh}
\pi_{nh} (x) = n^{-1} \sum_{i=1}^{n} \frac{1}{h} K \left(\frac{x - X_{t_i}}{h}\right),
\end{equation}
with $K(\cdot)$ the kernel density function and $h$ the bandwidth or smoothing parameter (see \citealp{rosenblatt1956remarks} and \citealp{parzen1962estimation}). Therefore, the hypotheses in~\eqref{eq:hypDistr} can be written as the null and alternative hypotheses
\begin{equation*}
\begin{cases}
\mathcal{H}_0 \colon \exists \boldsymbol{\theta}_0 \in \Theta \, \big/\, \pi(\cdot, \boldsymbol{\theta}_0) = \pi_0(\cdot), \\
\mathcal{H}_\text{a} \colon \pi_0(\cdot) \notin \Pi_{\boldsymbol{\theta}},
\end{cases}
\end{equation*}
with $\pi_0(\cdot)$ the true marginal density of the process.

When analyzing continuous time models, we can choose to use all the continuous sample path \citep{kutoyants2010goodness} or use a sampling scheme, as in~\eqref{eq:SDEdisc}. Regarding the later, \cite{Ait-Sahalia1996} and \cite{gao2004adaptive} proposed a test to check the specification of both drift and diffusion functions by comparing the parametric stationary density with the same density estimated nonparametrically, while the \cite{bickel1973some} test --based on the integrated squared error of the true and kernel density estimate-- was extended by \cite{lee2006bickel} to test the specification of the drift function. Under certain regularity conditions, the test statistics are asymptotically normal. However, in small samples, critical values computed based on the asymptotic distribution could lead to over reject the null hypothesis (see the discussion in Section~\ref{sec:boot}). Furthermore, the stationary marginal density does not characterize the full process, so tests based on it would not detect alternatives with the same marginal density as the null model. Thus, other authors, such as \cite{hong2004nonparametric} and \cite{ait2009nonparametric}, used the transition density, while the cumulative distribution function was considered in \cite{corradi2005bootstrap}. 
The former authors established the asymptotic null distributions of the proposed test statistics and the later used a block bootstrap procedure to calibrate the test distribution. The conditional characteristic function can also capture the full dynamic of the process, as the Fourier transform of the transitional density. This is addressed in \cite{chen2010characteristic}, where the authors constructed a nonparametric specification test for continuous-time Markov models using the conditional characteristic function. The test statistics proposed are asymptotically normal but the authors noted that using parametric bootstrap improved the finite-sample performance, as also stated in \cite{ait2009nonparametric}.

\subsection{Test based on the estimation of the regression function}

The GoF ideas from the later section can be extended to regression models. We can write the diffusion model in~\eqref{eq:SDEdisc} as
\begin{equation} \label{eq:SDEreg}
Y_{t_i} = m(X_{t_i}) + \sigma (X_{t_i}) \Delta^{-1/2} \varepsilon_{t_i},
\end{equation}
where $Y_{t_i} = (X_{t_{i+1}} - X_{t_i}) / \Delta$ and $\sqrt{\Delta}\varepsilon_{t_i} = (W_{t_{i+1}} - W_{t_i})$, therefore $\varepsilon_{t_i} \in N(0,1)$ for $i=1,\dots,n$. The goal is to test the specification of the drift
\begin{equation} \label{eq:mH0}
\mathcal{H}_0 \colon m \in \lbrace m(\cdot, \boldsymbol{\theta}) \colon \boldsymbol{\theta} \in \Theta \rbrace
\end{equation}
and the diffusion function,
\begin{equation*}
\mathcal{H}_0 \colon \sigma \in \lbrace \sigma(\cdot, \boldsymbol{\theta}) \colon \boldsymbol{\theta} \in \Theta \rbrace.
\end{equation*}
In the classical regression context, the target functions are usually the regression function of $Y_{t_i}$ over $X_{t_i}$, that is $m(x) = \mathbb{E} \left[ Y_{t_i} \mid X_{t_i} = x \right]$, and the conditional variance $\sigma^2 (x) = \mathbb{V}\text{ar} \left[ Y_{t_i} \mid X_{t_i} = x \right]$. 


\subsubsection{Smoothing-based test} 

The drift and diffusion terms can be estimated nonparametrically. Though different smoothing methods have been proposed, a large body of literature is focused in kernel type estimators (see \citealp{nadaraya1964estimating} and \citealp{watson1964smooth}). A kernel estimator for the drift and diffusion function (see \citealp{florens1993estimating}) is given by
\begin{equation} \label{eq:sigma_nh}
m_{nh} (x) = \sum_{i=1}^{n} W_{ni}(x) (X_{t_{i+1}} - X_{t_i})   \qquad \text{and} \qquad
\sigma^2_{nh} (x) = \sum_{i=1}^{n} W_{ni}(x) (X_{t_{i+1}} - X_{t_i})^2,
\end{equation}
with
\begin{equation*}
W_{ni}(x) = \frac{K \left(\frac{x - X_{t_i}}{h}\right)}{\Delta \sum_{j=1}^{n} K \left( \frac{x - X_{t_j}}{h} \right) }
\end{equation*}
where $\Delta \to 0$ is required to obtain consistency and asymptotic normality. \cite{florens1993estimating} considered a uniform kernel $K$, later \cite{jiang1997nonparametric} extended the results to the Gaussian kernel and \cite{bandi2003fully} to general stationary process and kernels.
The ideas of the previous section can be extended to the regression context, where an empirical process for the regression problem can be expressed as
\begin{equation} \label{eq:R1n}
R_{1n} (x) = \sqrt{nh} \big( m_{nh} (x) - \mathbb{E}_{\hat{\boldsymbol{\theta}}} \left[ m_{nh}(x) \right] \big) 
= \sqrt{nh} \sum_{i=1}^{n} W_{ni} (x) \left(Y_{t_i} - m_{\hat{\boldsymbol{\theta}}} (X_{t_i})\right),
\end{equation}
for the drift function (that be interpreted as a smoothing over the residuals $\hat{\eta}_{t_i} = Y_{t_i} - m_{\hat{\boldsymbol{\theta}}} (X_{t_i})$), which can easily be extended to the diffusion function. To implement the test, a root-$n$ consistent estimator $\hat{\boldsymbol{\theta}}$ of the true parameter $\boldsymbol{\theta}_0$ is needed, that is $\lVert \hat{\boldsymbol{\theta}} - \boldsymbol{\theta}_0 \rVert =\mathcal{O}_p (n^{-1/2})$, providing $\mathbb{E}_{\hat{\boldsymbol{\theta}}}$ an estimate of $\mathbb{E}_{\boldsymbol{\theta}_0}$ (see \citealp{lopez2021parametric}, for a review about parametric estimation methods for diffusion models). Therefore, a general test based on the empirical process $R_{1n}$ can be constructed by considering a continuous functional,
\begin{equation*}
T_n = \int R_{1n}^2 (x) \omega(x) \dif x.
\end{equation*}
For example, denoting $\eta_{t_i0} = Y_{t_i} - m_{\boldsymbol{\theta}_0} (X_{t_i})$, under the null hypothesis, since $\mathbb{E} \left[ \eta_{t_i0} \mid X_{t_i} \right] = 0$, we have
\begin{equation} \label{eq:testZheng}
\mathbb{E} \, \Big[ 
\eta_{t_i0} \;
\mathbb{E} \left[ \eta_{t_i0} \mid X_{t_i} \right] \;
\pi (X_{t_i})
\Big] = 0,
\end{equation}
and using a kernel-based sample analogue of~\eqref{eq:testZheng},
\begin{equation*}
T_{1n} = \frac{1}{n(n-1)} \sum_{i = 1}^n \sum_{\substack{j = 1 \\j \neq i}}^n h^{-1} K \left(\frac{X_{t_i} - X_{t_j}}{h} \right) \hat{\eta}_{t_i} \hat{\eta}_{t_j},
\end{equation*}
with $\hat{\eta}_{t_i} = \left(Y_{t_i} - m_{\hat{\boldsymbol{\theta}}} (X_{t_i}) \right)$, we obtain the test statistic introduced in~\cite{zheng1996consistent}. Note that in order to apply $U$-statistic theory, the negligible terms with $i=j$ are dropped. This idea was extended to diffusion models by \cite{gao2008specification}, where a nonparametric test for the specification of both drift and diffusion functions was developed for a discretized semiparametric diffusion model. \cite{arapis2006empirical} and \cite{masuda2011goodness} proposed nonparametric tests for the drift coefficient, while \cite{li2007testing} and \cite{chen2019nonparametric} proposals focused on the diffusion function. \cite{zheng2009testing} developed a nonparametric test for heteroscedasticity, build on the work of \cite{zheng1996consistent}. Regarding the test statistic distribution, asymptotic normality is the limiting distribution of the tests under $\mathcal{H}_0$. However, the critical values are usually obtained using simulation or resampling methods, as with finite samples the convergence is slow, a problem that arises in virtually all smoothing-based tests.


\subsubsection{Empirical regression process-based test} \label{sec:emp_reg}

Another regression-based approach is inspired by GoF test for distribution, were the selection of a smoothness parameter is avoided. This alternative methodology is based on the empirical estimator of the integrated regression function \citep{stute1997nonparametric}, $I(x) = \mathbb{E} \left[ Y \mathbbm{1}_{\lbrace X \leq x \rbrace} \right] = \int_{-\infty}^{x} m(y) \dif F_X(y)$, where $F_X$ is the marginal distribution function of $X$. 
An empirical estimator of the integrated regression is given by $I_n (x) = n^{-1} \sum_{i = 1}^n Y_{t_i} \mathbbm{1}_{\lbrace X_{t_i} \leq x \rbrace}$ and comparing the estimator $I_n(x)$ with its estimated values under the null hypothesis, a new empirical process can be constructed,
\begin{equation} \label{eq:R2n}
R_{2n} (x) = \sqrt{n} \left(I_n(x) - \mathbb{E}_{\hat{\boldsymbol{\theta}}} \left[ I_n(x) \right] \right) =  \frac{1}{\sqrt{n}} \sum_{i=1}^n \left(Y_{t_i} - m_{\hat{\boldsymbol{\theta}}} (X_{t_i}) \right) \; \mathbbm{1}_{\lbrace X_{t_i} \leq x \rbrace},
\end{equation}
The process $R_{2n}$ is a marked empirical process, where the marks are given by the residuals $\left(Y_{t_i} - m_{\hat{\boldsymbol{\theta}}} (X_{t_i}) \right)$. The test statistics $T_n = \Psi(R_{2n})$ can be constructed based on a distance of the resulting residual marked empirical process $R_{2n}$ in~\eqref{eq:R2n} from the expected zero mean, measured by Cramér-von Mises and Kolmogorov-Smirnov functionals $\Psi(\cdot)$. \cite{monsalve2011goodness} proposed a test for the drift and diffusion functions based on a marked empirical process of the residuals, constructed using estimators of the integrated regression function and the integrated conditional variance function, respectively. The distribution of the processes are approximated using bootstrap techniques, as tests based on an appropriately scaled discrepancy of the empirical process $R_{2n}$ in~\eqref{eq:R2n} are not generally asymptotically distribution free. In \cite{koul1999nonparametric} a transformation of $R_{2n}$ is considered in order to construct a process whose limiting distribution is known, so that tests based on it are asymptotically distribution free. To achieve this, a martingale transformation of the underlying process \citep{khmaladze1981martingale} is considered, and this idea was extended to diffusion models by several authors. An approach based on the innovation martingale to test the drift function was introduced in \citeauthor{negri2009goodness} (\citeyear{negri2009goodness}, \citeyear{negri2010goodness}) on the basis of the continuous observation and, subsequently, in \cite{masuda2011goodness} the method was extended to the case of discrete-time observations, obtaining an asymptotically distribution free test, as they proved that the limit distribution is the supremum of the standard Brownian motion. By means of the \cite{khmaladze1981martingale} martingale transformation of the marked empirical process, \cite{chen2015asymptotically} extends the proposal of \cite{monsalve2011goodness} to obtain a limiting distribution-free test for the parametric diffusion function. 

Another methodology that avoids the use of smoothing techniques, also based on cumulative sum processes, relies on the integrated variance function to test the parametric form of the variance function $\sigma^2(\cdot)$. 
\cite{dette2003test} used the integrated variance function for this purpose, also obtaining a homoscedasticity test, and proved that under the null hypothesis the test statistic had an asymptotic normal distribution. Subsequently, \cite{dette2006estimation} extended the latter proposal to volatility functions that depend on the variable $X_t$. \cite{dette2008testing} developed a test based on stochastic processes of the integrated volatility, proving its weak convergence to centered processes with Gaussian conditional distributions, and \cite{podolskij2008range} constructed a test statistic with a range-based estimation of the integrated volatility and the integrated quarticity, which converges weakly to a mixed Gaussian distribution.

\subsection{Generalized likelihood ratio test}

The generalized likelihood statistics were introduced in \cite{fan2001generalized} to overcome the drawbacks of nonparametric maximum likelihood ratio statistics. Considering the regression model in~\eqref{eq:SDEreg} 
and the conditional log-likelihood function $\ell (\cdot)$, the Generalized likelihood ratio test (GLRT) can be build as
\begin{equation*}
\Lambda_n = \ell_n (\mathcal{H}_\text{a}) - \ell_n (\mathcal{H}_0)  = \ell_n (m_{nh}) - \ell_n (m_{\hat{\boldsymbol{\theta}}}),
\end{equation*}
under the assumption that the error process is normally distributed. The log-likelihood under the null hypothesis is given by a maximum likelihood estimator and $\ell_n (\mathcal{H}_\text{a})$ is replaced by a generalized maximum likelihood estimator in the nonparametric context. Define the residual sum of squares under the null, $\text{RSS}_0 = n^{-1} \sum_{i=1}^{n} \left(Y_{t_i} - m_{\hat{\boldsymbol{\theta}}} (X_{t_i}) \right)^2$, and the alternative, $\text{RSS}_1 = n^{-1} \sum_{i=1}^{n} \left(Y_{t_i} - m_{nh} (X_{t_i}) \right)^2$, hypotheses, the statistic for the GLRT is given by
\begin{equation*}
\Lambda_n = \frac{n}{2} \log \frac{\text{RSS}_0}{\text{RSS}_1} \approx \frac{n}{2} \frac{\text{RSS}_0 - \text{RSS}_1}{\text{RSS}_1},
\end{equation*}
resembling the $F$-test construction for regression models. The test will reject the null hypothesis for large values of $\Lambda_n$, as under $\mathcal{H}_0$ the values of $\text{RSS}_0$ and $\text{RSS}_1$ will be close, whereas under the alternative hypothesis $\text{RSS}_0$ should become systematically larger than $\text{RSS}_1$.

The GLRT ideas developed in \cite{fan2001generalized} were extended to time-homogeneous diffusion models in \cite{fan2003reexamination}, while \cite{fan2003time} proposal was based on a generalized pseudo-likelihood ratio test and \cite{ait2009nonparametric} developed a specification test for continuous-time jump-diffusion process and derived the asymptotic null distribution. The asymptotic distribution of the GLRT exhibits the Wilks phenomenon, that is, the limit distribution of $\Lambda_n$ is asymptotically $\chi^2$, and a similar Wilks type of result holds under $\mathcal{H}_0$ in \cite{fan2003reexamination} test. \cite{fan2003time} used bootstrap methods to approximate the null distribution of the test statistic, as well as \cite{ait2009nonparametric}, though they also established its asymptotic null distribution.

\subsection{Empirical likelihood}

The empirical likelihood (EL) technique \citep{owen1988empirical} allows the construction of nonparametric likelihood for a parameter of interest. The idea of the EL is to maximize an objective function defined by a product of probability weights allocated to observations that, under certain constraints, characterize the functional curve to be tested. There exist a diversity of test statistic formulations via the EL, among them the local EL ratio test is a wide framework. The local version of the EL is constructed as
\begin{equation*}
\Lambda_n^L = -2 \int \log \left( \mathcal{L}_n \big( \tilde{m}(x, \hat{\boldsymbol{\theta}})n^n \big) \right) \omega (x) \dif x,
\end{equation*}
where $\mathcal{L}_n \left( \tilde{m}(x, \hat{\boldsymbol{\theta}}) \right) = \max \prod_{i=1}^{n} q_i(x)$, subject to $\sum_{i=1}^{n} q_i (x) = 1$ and
\begin{equation*}
\sum_{i=1}^{n} q_i (x) K_h (x - X_{t_i}) \left(Y_{t_i} - \tilde{m}(x, \hat{\boldsymbol{\theta}}) \right) = 0,
\end{equation*}
with $\tilde{m}(x, \hat{\boldsymbol{\theta}}) = \mathbb{E}_{\hat{\boldsymbol{\theta}}} \left[ m_{nh} (x) \right]$ the empirical local likelihood under the null hypothesis. This methodology has been extended to diffusion models, namely \cite{chen2008test} developed a EL test to parametrically specify the transitional density of the process based on kernel estimation. They used the EL to formulate a statistic for each bandwidth that is a Studentized $L_2$-distance between the nonparametric and parametric estimator of the transitional density implied by the parametric process, where a parametric bootstrap procedure was used to obtain the critical values, due to the slow convergence of the asymptotic distribution. \cite{chen2011simultaneous} proposed a nonparametric test for the parametric specification of both the conditional mean and variance functions based on an EL statistic, calibrated using bootstrap techniques.

\subsection{Test distribution and calibration} \label{sec:boot}

The null hypothesis $\mathcal{H}_0$ 
is rejected at the level $\alpha$ if the test statistic $T_n$ exceeds the $\alpha$-level critical value $c_{1-\alpha}$, that is $T_n \geq c_{1-\alpha}$, where $c_{1-\alpha}$ satisfies
\begin{equation*}
\mathbb{P} \left( T_n \geq c_{1-\alpha} \right) = \alpha.
\end{equation*}
If the distribution of the test statistic is known, the value $c_{1-\alpha}$ can be determined. Alternatively, the distribution can be approximated by Monte Carlo procedures or resampling methods, such as bootstrap techniques. Though the limiting distribution of the test statistics is available for some procedures, asymptotic critical values may not be accurate in practice, in particular when the sample size is not sufficiently large and the convergence rate to the limit distribution is slow. The intricacies of empirical interest rate data, highly persistent, can further emphasize this, as the near-nonstationarity of these series can lead to over rejection of the null hypothesis when computing critical values based on the asymptotic distribution derived under stationarity. This was noted in~\cite{pritsker1998nonparametric}, who studied the small sample behavior of the asymptotically normally distributed test of \cite{Ait-Sahalia1996}. Short term rates exhibit very low speeds of mean reversion and, despite the process is still formally stationary, the asymptotics based on assuming stationarity are slow to converge, to such an extent that \cite{ait2012stationarity} investigated that it may well be that the asymptotic behavior derived under nonstationarity is a better approximation to the finite sample behavior of the test. Therefore, when the data generating process is stationary but exhibits high persistence, the ability of the limiting distribution to approximate the finite sample behavior of the test deteriorates. As an alternative, the martingale transformation can be used to obtain asymptotically distribution free tests (see, \citealp{khmaladze1979omega}; \citealp{khmaladze1981martingale}; \citealp{stute1998model}; \citealp{khmaladze2004martingale}). These transformation has been used in the context of diffusion models, but mainly focused on the market empirical process of the residuals (see Section~\ref{sec:emp_reg}).
On the other hand, bootstrap methods can be used to calibrate the distribution of the test statistic $T_n$ under the null hypothesis, where the critical value $c_{1-\alpha}$ is approximated by $c_{1-\alpha}^*$ such that
\begin{equation*}
\mathbb{P}^* \left( T_n^* \geq c_{1-\alpha}^* \right) = \alpha,
\end{equation*}
with $\mathbb{P}^*$ the probability measure generated by the bootstrap sample and $c_{1-\alpha}^*$ can be obtained by means of Monte Carlo techniques with a statistic of order $\lceil B(1-\alpha)\rceil$ from $B$ bootstrap replicates $\lbrace T_n^{*j} \rbrace_{j=1}^B$, that is, $c_{1-\alpha}^* = T_n^{*\lceil B(1-\alpha)\rceil}$. Parametric bootstrap procedures were used in different works (i.e., \citealp{gao2008specification}; \citealp{ait2009nonparametric}; \citealp{chen2010characteristic}; \citealp{monsalve2011goodness}), which can be implemented using the following simulation procedure:
\begin{enumerate}
	\item Generate the bootstrap sample $\left\lbrace \left(X_{t_i}^*, Y_{t_i}^*\right) \right\rbrace_{i=1}^n$, with $Y_{t_i}^* = m(X_{t_i}, \hat{\boldsymbol{\theta}}) + \sigma (X_{t_i}, \hat{\boldsymbol{\theta}}) \Delta^{-1/2} \varepsilon_{t_i}^*$, considering a fixed design ($X_{t_i}^* \coloneqq X_{t_i}$) and where $\lbrace \varepsilon_{t_i}^* \rbrace$ are i.i.d. random variables following a standard Gaussian distribution, independent of $\lbrace X_{t_i} \rbrace$.
	
	\item Estimate $\hat{\boldsymbol{\theta}}^*$ from the bootstrap sample $\left\lbrace \left(X_{t_i}, Y_{t_i}^*\right) \right\rbrace_{i=1}^n$ obtained in the previous step.
	
	\item Define the bootstrap statistic $T_n^*$ to be the version of $T_n$ with $\lbrace \left(X_{t_i}, Y_{t_i}\right) \rbrace_{i=1}^n$ and $\hat{\boldsymbol{\theta}}$ being replaced by $\left\lbrace \left(X_{t_i}, Y_{t_i}^*\right) \right\rbrace_{i=1}^n$ and $\hat{\boldsymbol{\theta}}^*$.
	
	\item Repeat $B$ times the above steps, obtaining the bootstrap replicates $\lbrace T_n^{*j} \rbrace_{j=1}^B$.
	
	\item Approximate the critical value, $c_{1-\alpha}^* = T_n^{*\lceil B(1-\alpha)\rceil}$, or the $p$-value, $\# \lbrace T_n^{*j} > T_n \rbrace / B$.
\end{enumerate}

%
%
%

\subsection{Overcoming the curse of dimensionality} 

Some of the tests introduced in this section are subject of the curse of dimensionality, which refers to poor performances of local smoothing methods for multivariate data \citep{lavergne2008breaking}, as the behavior of nonparametric estimators quickly deteriorates as dimension increases, due to the sparsity of data in multidimensional spaces \citep{stone1980optimal}. Test statistics built by the comparison of a nonparametric estimator of the model and an estimator under the null hypothesis (based on the $R_{1n}$ process in~\eqref{eq:R1n}) or by the comparison of the empirical estimator of the integrated regression function with its estimated values under $\mathcal{H}_0$ (based on the $R_{2n}$ process in~\eqref{eq:R2n}) are subject of the curse of dimensionality as the dimension increases. The effect of the increasing dimension deteriorates the power of the tests. This problem has been addressed by several authors, regarding tests based on smoothing techniques (that is, on the process $R_{1n}$), see \cite{lavergne2008breaking} and \cite{xia2009model}, whose proposals were inspired on the projection pursuit ideas. For the test based on the empirical regression process, in \cite{stute2008model} the process $R_{2n}$ was replaced by a residual empirical process marked by proper functions of the regressors. \cite{stute2006model} addressed this problem in higher order time series and \cite{escanciano2006consistent} overcame the curse of dimensionality by using a residual marked empirical process based on projections.

Though different adaptations have been proposed in order to avoid the curse of dimensionality, few have been studied for diffusion processes. In the next section, we use the ideas of correlation distance in \cite{szekely2007measuring} to test the parametric specification of diffusion processes, which circumvents the curse of dimensionality problem.





%

\section{Testing parametric specification with correlation distance} \label{sec:3}


%


Given the discretized diffusion model in~\eqref{eq:SDEreg},
\begin{equation*}
Y_{t_i} = m(X_{t_i}) + \frac{\sigma (X_{t_i})}{\sqrt{\Delta}} \varepsilon_{t_i},
\end{equation*}
where $Y_{t_i} = (X_{t_{i+1}} - X_{t_i}) / \Delta$ and $\varepsilon_{t_i} \in N(0,1)$ for $i=1,\dots,n$, we work with a fixed time span with $T=1$, that is, the unit interval $[0,1]$ and sampling interval $\Delta = 1/n$. Thus, as the sample size $n$ increases, the time span remains fixed and $\Delta \to 0$ as $n\to \infty$. Under stationarity and ergodicity, we have that $\lbrace \varepsilon_{t_i} \rbrace_{i=1}^n$,
\begin{equation*}
\varepsilon_{t_i} = \frac{Y_{t_i} - m(X_{t_{i}})}{\sigma(X_{t_i})/\sqrt{\Delta}},
\end{equation*}
are an i.i.d. sequence independent of the stochastic process $\lbrace X_{t_i} \rbrace_{i=1}^n$ with $\mathbb{E} \left[ \varepsilon_{t_i} \mathrel{\big|} \mathcal{F}_t \right] = 0$. Therefore, the parametric specification of the drift and diffusion terms in~\eqref{eq:SDEreg} can be checked by testing the independence of the variables $\lbrace \varepsilon_{t_i}, X_{t_i} \rbrace_{i=1}^n$. 

Based on the distribution of the residuals, we construct a test for the parametric specification of the diffusion model motivated by \cite{szekely2007measuring}. Let  $\varphi_{\varepsilon,X} (t,s) = \mathbb{E} \left[ e^{i(\langle t, \varepsilon \rangle + \langle s, X \rangle)} \right]$ denote the joint characteristic function of $X_{t_i} \in \mathbb{R}$ and $\varepsilon_{t_i} \in \mathbb{R}$ and $\varphi_\varepsilon (t) = \varphi_{\varepsilon,X} (t,0) $ and $\varphi_X (s) = \varphi_{\varepsilon,X} (0,s) $ the characteristic functions of $\varepsilon_{t_i}$ and $X_{t_i}$, respectively. In terms of characteristic functions, $X_{t_i}$ and $\varepsilon_{t_i}$ are independent if and only if $\varphi_{\varepsilon,X} = \varphi_\varepsilon \varphi_X$. Therefore, to test the hypothesis of independence
\begin{equation} \label{eq:H0char}
\mathcal{H}_0 \colon \varphi_{\varepsilon,X} = \varphi_\varepsilon \varphi_X \qquad \text{ vs. } \qquad
\mathcal{H}_\text{a} \colon \varphi_{\varepsilon,X} \neq \varphi_\varepsilon \varphi_X,
\end{equation}
we can use a distance dependence measure such as the distance covariance \citep{szekely2007measuring} defined as
\begin{align} \label{eq:dcov}
\mathcal{V}^2 (\varepsilon_t,X_t;\mu) &\coloneqq \left\lVert \varphi_{\varepsilon,X} (t,s) - \varphi_\varepsilon (t) \varphi_X (s) \right\rVert^2_\mu \\
&= 
\int_{\mathbb{R}} \int_{\mathbb{R}} \left\lvert \varphi_{\varepsilon,X} (t,s) - \varphi_\varepsilon (t) \varphi_X (s) \right\rvert^2 \mu(t,s) \dif t \dif s,
\end{align}
where $\mu (t,s) = \left( c^2 \lvert t \rvert^{\alpha +1} \lvert s \rvert^{\alpha +1} \right)^{-1}$ is a non-negative weight function used in the $\left\lVert \cdot \right\rVert_\mu$-norm, for a constant $c>0$ and $\alpha \in (0,2)$. Following \cite{szekely2007measuring}, we fix $\alpha = 1$ and thus assume $\mu (t,s) = (c^2 t^2 s^2)^{-1}$, with $c=\pi$, as with this choice of $\mu$ the distance correlation, $\mathcal{R}_\mu = \mathcal{V} (\varepsilon_t,X_t;\mu) / (\mathcal{V} (\varepsilon_t,\varepsilon_t;\mu)\mathcal{V} (X_t,X_t;\mu))^{1/2}$, is invariant relative to scale and orthogonal transformations. 

As $\mathcal{V}^2 (\varepsilon_t,X_t;\mu) = 0$ if and only if $\varepsilon_{t}$ and $X_{t}$ are independent, that is, $\varphi_{\varepsilon,X} = \varphi_\varepsilon \varphi_X$, a test statistic can be constructed based in the distance covariance. Under the null hypothesis we have $\mathcal{V}_0^2 (\varepsilon_{t_i0},X_{t_i};\mu) = 0$, with $\varepsilon_{t_i0} = (Y_{t_i} - m_0(X_{t_{i}}))/(\sigma_0(X_{t_i})/\sqrt{\Delta})$, since $\mathbb{E} \left[ \varepsilon_{t_i0} \mathrel{\big|} \mathcal{F}_t \right] = 0$. 
To implement the test we need to consider a root-$n$ consistent estimator $\hat{\boldsymbol{\theta}}$ of the true parameter $\boldsymbol{\theta}_0$, obtaining the residuals $\hat{\varepsilon}_{t_i} = (Y_{t_i} - m_{\hat{\boldsymbol{\theta}}}(X_{t_{i}}))/(\sigma_{\hat{\boldsymbol{\theta}}}(X_{t_i})/\sqrt{\Delta})$. 
An empirical estimator of the distance covariance~\eqref{eq:dcov} can be obtained by replacing the characteristic functions $\varphi$ by their sample analogs
\begin{align*}
\mathcal{V}^2_n (\hat{\varepsilon}_t,X_t;\mu) &= \left\lVert \varphi^{(n)}_{\hat{\varepsilon},X} (t,s) - \varphi^{(n)}_{\hat{\varepsilon}} (t) \varphi^{(n)}_X (s) \right\rVert^2_\mu \\
&= 
\int_{\mathbb{R}} \int_{\mathbb{R}} \left\lvert \varphi^{(n)}_{\hat{\varepsilon},X} (t,s) - \varphi^{(n)}_{\hat{\varepsilon}} (t) \varphi^{(n)}_X (s) \right\rvert^2 \mu(t,s) \dif t \dif s,
\end{align*}
where the empirical characteristic functions are given by $\varphi^{(n)}_{\varepsilon,X} (t,s) = n^{-1} \sum_{j=1}^n e^{i\langle t, \varepsilon_{t_j} \rangle + i\langle s, X_{t_j} \rangle}$, $\varphi^{(n)}_\varepsilon (t) = \varphi^{(n)}_{\varepsilon,X} (t,0) $ and $\varphi^{(n)}_X (s) = \varphi^{(n)}_{\varepsilon,X} (0,s) $. However, due to the computational cost of this definition, a simpler formula with practical advantages is preferable. An equivalent (see \citealp{szekely2007measuring}, Theorem 1) empirical distance covariance is defined by
\begin{equation*}
\mathcal{V}_n^2 (\hat{\varepsilon}_{t_i},X_{t_i}) = \frac{1}{n^2} \sum_{k,l=1}^{n} A_{kl} B_{kl},
\end{equation*}
where $A_{kl}$ and $B_{kl}$ are linear functions of the pairwise distances between sample elements, 
\begin{equation*}
A_{kl} = a_{kl} - \bar{a}_{k\cdot} - \bar{a}_{\cdot l} + \bar{a}_{\cdot\cdot}, \qquad 
B_{kl} = b_{kl} - \bar{b}_{k\cdot} - \bar{b}_{\cdot l} + \bar{b}_{\cdot\cdot},
\end{equation*}
with $a_{kl} = \lvert \hat{\varepsilon}_{t_k} - \hat{\varepsilon}_{t_l} \rvert$ and $b_{kl} = \lvert X_{t_k} - X_{t_l} \rvert$ the Euclidean distance matrices and
\begin{equation*}
\bar{a}_{k\cdot} = \frac{1}{n} \sum_{l=1}^n a_{kl}, \qquad
\bar{a}_{\cdot l} = \frac{1}{n} \sum_{k=1}^n a_{kl}, \qquad
\bar{a}_{\cdot\cdot} = \frac{1}{n^2} \sum_{k,l=1}^n a_{kl},
\end{equation*}
for $k,l = 1, \dots,n$ and, similarly, define $\bar{b}_{k\cdot}, \bar{b}_{\cdot l}$ and $\bar{b}_{\cdot\cdot}$. Given the empirical distance covariance $\mathcal{V}_n^2$, the test of independence for the null hypothesis in~\eqref{eq:H0char} is based on the statistic 
\begin{equation*}
T_n = n \frac{\mathcal{V}_n^2 (\hat{\varepsilon}_{t_i},X_{t_i})}{S_2},
\end{equation*}
%
%
%
where $S_2 = \bar{a}_{\cdot\cdot} \bar{b}_{\cdot\cdot}$. The test rejects independence for large values of $T_n$ and a resampling procedure can be used to determine the quantiles of the test statistic, such as the parametric bootstrap procedure in Section~\ref{sec:boot}. In \cite{szekely2007measuring}, replicates of $T_n$ were computed under random permutations of one of the sample indices. 

\section{Simulation study} \label{sec:4}

In this Section, we carry out a simulation study to compare the performance of the methods described, namely, the empirical regression (henceforth abbreviated as ER) process-based test of \cite{monsalve2011goodness} and its asymptotically distribution-free (ADF) transformation \citep{chen2015asymptotically}, for both Kolmogorov-Smirnov (KS) and Cramér-von Mises (CvM) test statistics; the nonparametric (NP) proposal of \cite{gao2008specification}; and the test based on distance covariance (DCOV) introduced in Section~\ref{sec:3}.

The finite sample behavior of tests for higher order time series is also addressed by studying the effect of the curse of dimensionality in the test based on the process $R_{2n}$ in~\eqref{eq:R2n}, as \cite{monsalve2011goodness} proposal, and the distance covariance test of Section~\ref{sec:3}, designed for avoiding the curse of dimensionality.


\subsection{Comparative study}

We study the finite sample properties of the tests using Monte Carlo simulations, given the diffusion model in~\eqref{eq:SDEparam},
\begin{equation*} 
\dif X_t = m(X_t, \boldsymbol{\theta})\dif t + \sigma(X_t, \boldsymbol{\theta}) \dif W_t.
\end{equation*}
The sample paths for the models are simulated using Milsteins scheme \citep{mil1979method} and we consider the time interval $[0,1]$, with $\Delta = 1 / n$, generating $1\,000$ random samples with sample sizes $n \in \lbrace 100, 250, 500, 1\,000 \rbrace$. The first thousand observations are discarded to lessen the impact of the initial value. Regarding the methods that require bootstrap resamples to calibrate the test, we chose $B = 1\, 000$ replicates. The unknown parameters are estimated using the \cite{kalman1960new} filter.

As null hypothesis we test the parametric specification of the diffusion function, $\mathcal{H}_0\colon \sigma^2 (x) = \sigma^2 x^2$, where $\sigma$ is an unknown parameter, and simulate under different scenarios (similar design as in \citealp{dette2008testing}):

\begin{itemize}
	\item Size simulation: we consider five scenarios with diffusion function ${\sigma^2(x) = x^2}$, therefore $\sigma=1$, and drift functions $m_1(X_t)=0$, $m_2(X_t)=2$, $m_3(X_t)=X_t$, $m_4(X_t)=2-X_t$ and $m_5(X_t,t)=tX_t$.

	\item Power simulation: we consider five scenarios with drift function ${m(X_t)=2-X_t}$ and diffusion functions $\sigma_1^2(X_t)=1+X_t^2$, $\sigma_2^2(X_t)=1$, $\sigma_3^2(X_t)=5\lvert X_t\rvert^{1.5}$, $\sigma_4^2(X_t)=5\lvert X_t\rvert$ and $\sigma_5^2(X_t)=(1+X_t)^2$.
\end{itemize}

Table~\ref{tab:comp_size} and~\ref{tab:comp_power} show the empirical rejection rates for $\alpha = 0.05$. Regarding the sizes in Table~\ref{tab:comp_size}, all methods approximate the nominal level $\alpha$, though the ADF test shows some under-rejection through the different scenarios, while the distance covariance test slightly over-rejects. Nevertheless, none of the methods show repeated miscalibration. With respect to power in Table~\ref{tab:comp_power}, all the methods perform closely, with the distance covariance proposal showing less power for the smallest sample size in certain scenarios, as $\sigma_3^2(X_t)=5\lvert X_t\rvert^{1.5}$ for $n=100$.

\begin{table}
	\centering
	\small
	\renewcommand{\arraystretch}{0.8}
	\begin{tabular}{crrrrrrr}
		\toprule
		&   & \multicolumn{2}{c}{ER} & \multicolumn{2}{c}{ADF} &  &  \\
		\cmidrule(l){3-4}\cmidrule(l){5-6} 
		$m(\cdot)$ & \multicolumn{1}{c}{n} & \multicolumn{1}{c}{KS} & \multicolumn{1}{c}{CvM} & \multicolumn{1}{c}{KS} & \multicolumn{1}{c}{CvM} & \multicolumn{1}{c}{NP} & \multicolumn{1}{c}{DCOV} \\
		\midrule
		\multirow{4}[2]{*}{$ 0 $} 
		& $100 $  & $0.038$ & $0.044$ & $0.043$ & $\mathbf{0.069}$ & $0.060$ & $\mathbf{0.065}$ \\
		& $250 $  & $0.048$ & $0.055$ & $0.049$ & $0.057$ & $0.059$ & $0.046$ \\
		& $500 $  & $0.052$ & $0.044$ & $0.052$ & $0.061$ & $0.046$ & $0.047$ \\
		& $1000$  & $0.039$ & $0.044$ & $0.041$ & $0.055$ & $0.047$ & $0.051$ \\
		\midrule
		\multirow{4}[2]{*}{$ 2 $} 
		& $100 $  & $0.057$ & $\mathbf{0.072}$ & $\mathbf{0.028}$ & $0.044$ & $0.052$ & $\mathbf{0.076}$ \\
		& $250 $  & $0.037$ & $0.040$ & $0.033$ & $0.038$ & $0.055$ & $0.062$ \\
		& $500 $  & $0.044$ & $0.041$ & $\mathbf{0.033}$ & $\mathbf{0.036}$ & $0.044$ & $0.064$ \\
		& $1000$  & $0.038$ & $0.037$ & $0.033$ & $0.036$ & $0.038$ & $0.049$ \\
		\midrule
		\multirow{4}[2]{*}{$ X_t $} 
		& $100 $  & $0.044$ & $0.058$ & $0.031$ & $0.050$ & $0.047$ & $0.060$ \\
		& $250 $  & $0.060$ & $0.063$ & $0.042$ & $0.054$ & $0.056$ & $0.056$ \\
		& $500 $  & $0.041$ & $0.048$ & $0.034$ & $0.036$ & $0.048$ & $0.053$ \\
		& $1000$  & $0.059$ & $0.061$ & $0.033$ & $0.044$ & $0.044$ & $0.062$ \\
		\midrule
		\multirow{4}[2]{*}{$ 2-X_t $} 
		& $100 $  & $0.043$ & $0.055$ & $\mathbf{0.032}$ & $0.048$ & $\mathbf{0.067}$ & $\mathbf{0.080}$ \\
		& $250 $  & $0.046$ & $0.053$ & $\mathbf{0.034}$ & $\mathbf{0.035}$ & $0.061$ & $0.059$ \\
		& $500 $  & $0.050$ & $0.053$ & $0.037$ & $0.037$ & $0.053$ & $0.057$ \\
		& $1000$  & $0.047$ & $0.062$ & $0.037$ & $0.046$ & $0.052$ & $0.053$ \\
		\midrule
		\multirow{4}[2]{*}{$ tX_t $} 
		& $100 $  & $0.058$ & $\mathbf{0.071}$ & $\mathbf{0.030}$ & $\mathbf{0.033}$ & $\mathbf{0.066}$ & $\mathbf{0.065}$ \\
		& $250 $  & $0.051$ & $0.056$ & $0.045$ & $0.054$ & $0.060$ & $\mathbf{0.068}$ \\
		& $500 $  & $0.058$ & $\mathbf{0.067}$ & $0.040$ & $0.042$ & $0.037$ & $0.057$ \\
		& $1000$  & $0.055$ & $0.055$ & $0.037$ & $0.039$ & $0.050$ & $0.048$ \\
		\bottomrule
	\end{tabular}%
	\caption{\label{tab:comp_size} Empirical sizes for $\alpha = 0.05$. Boldface indicates the rates of rejection outside a 95\% confidence interval for the nominal level $\alpha$.}%
\end{table}%

\begin{table}[H]
	\centering
	\small
	\renewcommand{\arraystretch}{0.8}
	\begin{tabular}{crrrrrrr}
		\toprule
		&   & \multicolumn{2}{c}{ER} & \multicolumn{2}{c}{ADF} &  &  \\
		\cmidrule(l){3-4}\cmidrule(l){5-6} 
		$\sigma^2(\cdot)$ & \multicolumn{1}{c}{n} & \multicolumn{1}{c}{KS} & \multicolumn{1}{c}{CvM} & \multicolumn{1}{c}{KS} & \multicolumn{1}{c}{CvM} & \multicolumn{1}{c}{NP} & \multicolumn{1}{c}{DCOV} \\
		\midrule
		\multirow{4}[2]{*}{$1 + X_t^2$} 
		& $100 $  & $0.605$ & $0.599$ & $0.496$ & $0.637$ & $0.580$ & $0.493$ \\
		& $250 $  & $0.867$ & $0.860$ & $0.803$ & $0.854$ & $0.800$ & $0.772$ \\
		& $500 $  & $0.976$ & $0.967$ & $0.938$ & $0.962$ & $0.955$ & $0.955$ \\
		& $1000$  & $0.997$ & $0.996$ & $0.970$ & $0.985$ & $0.990$ & $0.993$ \\
		\midrule
		\multirow{4}[2]{*}{$1$} 
		& $100 $  & $0.579$ & $0.582$ & $0.587$ & $0.742$ & $0.678$ & $0.592$ \\
		& $250 $  & $0.943$ & $0.953$ & $0.934$ & $0.965$ & $0.964$ & $0.942$ \\
		& $500 $  & $1.000$ & $0.999$ & $0.965$ & $0.979$ & $0.997$ & $0.995$ \\
		& $1000$  & $1.000$ & $1.000$ & $0.983$ & $0.989$ & $0.999$ & $0.996$ \\
		\midrule
		\multirow{4}[2]{*}{$5\,\lvert X_t\rvert^{1.5}$} 
		& $100 $  & $0.514$ & $0.551$ & $0.529$ & $0.735$ & $0.478$ & $0.235$ \\
		& $250 $  & $0.869$ & $0.873$ & $0.872$ & $0.940$ & $0.764$ & $0.564$ \\
		& $500 $  & $0.974$ & $0.973$ & $0.990$ & $0.995$ & $0.936$ & $0.893$ \\
		& $1000$  & $1.000$ & $1.000$ & $0.999$ & $1.000$ & $0.990$ & $0.994$ \\
		\midrule
		\multirow{4}[2]{*}{$5\,\lvert X_t\rvert$} 
		& $100 $  & $0.830$ & $0.810$ & $0.818$ & $0.909$ & $0.811$ & $0.809$ \\
		& $250 $  & $0.961$ & $0.960$ & $0.972$ & $0.993$ & $0.961$ & $0.984$ \\
		& $500 $  & $0.989$ & $0.989$ & $0.992$ & $0.998$ & $0.988$ & $1.000$ \\
		& $1000$  & $0.993$ & $0.993$ & $0.999$ & $1.000$ & $0.995$ & $1.000$ \\
		\midrule
		\multirow{4}[2]{*}{$\big(1 + X_t\big)^2$} 
		& $100 $  & $0.806$ & $0.790$ & $0.694$ & $0.799$ & $0.778$ & $0.654$ \\
		& $250 $  & $0.954$ & $0.950$ & $0.895$ & $0.934$ & $0.940$ & $0.893$ \\
		& $500 $  & $0.996$ & $0.992$ & $0.945$ & $0.968$ & $0.990$ & $0.980$ \\
		& $1000$  & $1.000$ & $0.999$ & $0.968$ & $0.980$ & $0.999$ & $0.991$ \\
		\bottomrule
	\end{tabular}%
	\caption{\label{tab:comp_power} Empirical power for $\alpha = 0.05$.}%
\end{table}%

\subsection{Higher order time series and the curse of dimensionality}

To illustrate the effect of the curse of dimensionality, we conduct a simulation study with higher order time series. A continuous-time analogue of the $p \in \mathbb{N}$ order AR($p$) process is the continuous-time Gaussian autorregresive process CAR($p$) symbolically defined to be a stationary solution of the SDE
\begin{equation*}
a(D) Y(t) = \sigma DW(t),
\end{equation*}
with $a(D) = D^p + \alpha_1 D^{p-1} + \dots + \alpha_p$, where the operator $D$ denotes differentiation with respect to $t$ and $\lbrace W(t) \rbrace_{t \geq 0}$ is a standard Wiener process. Since $DW(t)$ does not exist in the usual sense (it does not exist as a random function), a meaningful interpretation can be given by writing it in the state-space form,
\begin{equation*}
Y(t) = \mathbf{b}^\top \mathbf{X}(t),
\end{equation*}
$\mathbf{b}^\top = (1,0,\dots,0)$, where the state vector $\mathbf{X}(t) = \left(X^{(0)}(t), \dots, X^{(p-1)}(t) \right)^\top$, with $X^{(j)} (t)$ the $j^\text{th}$ mean-square and pathwise derivative $D^j X^{(0)} (t)$, satisfies the Itô equation
\begin{equation*}
\dif \mathbf{X} (t) = \mathbf{A} \mathbf{X} (t) \dif t + \sigma \mathbf{e} \dif W(t),
\end{equation*}
with $\sigma \in (0,\infty)$,
\begin{equation*}
\mathbf{A} = 
	\begin{bmatrix}
	0 & 1 & 0 & \cdots & 0 \\
	0 & 0 & 1 & \cdots & 0 \\
	\vdots & \vdots & \vdots & \ddots & \vdots \\
	0 & 0 & 0 & \cdots & 1 \\
	-\alpha_p & -\alpha_{p-1} & -\alpha_{p-2} & \cdots & -\alpha_1
	\end{bmatrix} 
\qquad \text{and} \qquad
\mathbf{e} =
	\begin{bmatrix}
	0 \\ 0 \\ \vdots \\ 0 \\ 1
	\end{bmatrix},
\end{equation*}
under the initial condition that $\mathbf{X}(0)$ is normally distributed and independent of $W(t)$ for all $t \geq 0$. The CAR($p$) process $\lbrace Y(t) \rbrace_{t \geq 0}$ exists and satisfies
\begin{equation*}
Y(t) = \mathbf{b}^\top e^{\mathbf{A}t} \mathbf{X}(0) + \sigma \mathbf{b}^\top \int_{0}^{t} e^{\mathbf{A}(t-u)} \mathbf{e} \dif W(u),
\end{equation*}
for $t \geq 0$. Note that the CAR($1$) is the Ornstein-Uhlenbeck process, usually parametrized as
\begin{equation} \label{eq:OU}
\dif Y_t = \kappa (\mu - Y_t) \dif t + \sigma \dif W_t, \qquad t \in [0,T],
\end{equation}
where $\mu$ is the long term mean, $\kappa$ the rate of mean reversion and $\sigma > 0$ the volatility around the mean.
%
%
To estimate the autoregressive coefficients $\alpha_1, \alpha_2, \dots, \alpha_p$, we proceed to estimate the analogue discrete-time ARMA process $\lbrace Y_n^{(\Delta)} \coloneqq Y(n\Delta), n=0,1,2,\dots \rbrace$ by maximum likelihood and from that estimates we obtain $\hat{\alpha}_1, \hat{\alpha}_2, \dots, \hat{\alpha}_p$.

Regarding non-linear time series, we consider the continuous-time analogue of the threshold models of \cite{tong1983threshold}, the continuous-time threshold autoregressive process CTAR($1$), given by
%
\begin{equation} \label{eq:CTAR}
\begin{aligned}
DY(t) + \alpha_1 Y(t) &= DW(t) + \alpha_1 \mu_1, &&\text{ if } Y(t) \leq r, \\
DY(t) + \alpha_2 Y(t) &= DW(t) + \alpha_2 \mu_2, &&\text{ if } Y(t) > r,
\end{aligned}
\end{equation}
with threshold $r$.

To study the behavior of the tests as the dimension $p$ increases we use Monte Carlo simulations. As null hypothesis we test the CAR($p$) process, for different order $p$, to illustrate the finite sample behavior regarding power and size:

\begin{itemize}
	\item Size simulation: we generate samples from the CAR($1$) process, namely the Ornstein-Uhlenbeck process in~\eqref{eq:OU}, noting that the CAR($1$) may be embedded in CAR($p$), and test for CAR($p$) with $p \in \lbrace 1,2,3,4,5 \rbrace$. The parameter values are $(\mu, \kappa, \sigma^2) = (0.08, 0.5, 0.25)$.
	
	\item Power simulation: we consider the CTAR($1$) process in~\eqref{eq:CTAR} with parameters $(\mu_1, \alpha_1, \mu_2, \alpha_2) = (4, 0.5, -1.25, -0.4)$ and $r = 1$, and test for CAR($p$) with $p \in \lbrace 1,2,3,4,5 \rbrace$.
\end{itemize}

Table~\ref{tab:curse_size} depicts the rate of rejection under the null hypothesis, where the empirical regression process-based proposal of \cite{monsalve2011goodness}, both for the Kolmogorov-Smirnov and the Cramér-von Mises statistics, and the covariance distance test of Section~\ref{sec:3} attained the nominal level $\alpha = 0.05$ under the null hypothesis. However, when analyzing the empirical power in Table~\ref{tab:curse_power}, the curse of dimensionality arises as the dimension $p$ increases for both the Kolmogorov-Smirnov and the Cramér-von Mises criteria, where power drastically decreases for higher order $p$. Except for dimension $p=1$ when both tests are equivalent and produce similar results, the covariance distance test is more powerful, circumventing the curse of dimensionality.

\begin{table}[H]
	\centering
	\adjustbox{max width=\textwidth}{
	\begin{tabular}{rrrrrrrrrrrrrrrr}
		&       & \multicolumn{4}{c}{Kolmogorov-Smirnov} &       & \multicolumn{4}{c}{Cramér-von Mises} & & \multicolumn{4}{c}{Covariance distance} \\
		\cmidrule{3-6}\cmidrule{8-11}\cmidrule{13-16}    \multicolumn{1}{c}{p} & n: & $ 100 $   & $ 250 $   & $ 500 $   & $ 1000 $  && $ 100 $   & $ 250 $   & $ 500 $   & $ 1000 $  && $ 100 $   & $ 250 $   & $ 500 $   & $ 1000 $ \\
		\midrule
		1 && $0.045$ & $0.054$ & $0.039$ & $0.057$ && $0.056$ & $0.049$ & $0.037$ & $0.053$ && $0.044$ & $0.041$ & $0.042$ & $0.048$ \\
		2 && $0.060$ & $0.055$ & $0.062$ & $0.053$ && $\mathbf{0.066}$ & $0.046$ & $0.056$ & $0.049$ && $0.052$ & $0.048$ & $0.041$ & $0.046$ \\
		3 && $0.052$ & $0.039$ & $0.044$ & $0.042$ && $0.051$ & $0.040$ & $0.039$ & $0.047$ && $0.050$ & $0.048$ & $0.042$ & $0.045$ \\
		4 && $0.045$ & $0.034$ & $0.055$ & $0.051$ && $0.039$ & $0.042$ & $0.050$ & $0.039$ && $0.049$ & $0.047$ & $0.039$ & $0.043$ \\
		5 && $0.047$ & $0.041$ & $0.053$ & $0.054$ && $0.042$ & $0.046$ & $0.054$ & $0.043$ && $0.054$ & $0.045$ & $0.038$ & $0.046$ \\
		\bottomrule
	\end{tabular}%
	}
	\caption{\label{tab:curse_size} Empirical levels under the null hypothesis, with $\alpha = 0.05$, testing the CAR($p$) process with order $p \in \lbrace 1,2,3,4,5 \rbrace$. Bolfaced the rates of rejection outside a 95\% confidence interval for the nominal level $\alpha$.}%
\end{table}%

\begin{table}[H]
	\centering
	\adjustbox{max width=\textwidth}{
	\begin{tabular}{rrrrrrrrrrrrrrrr}
		&       & \multicolumn{4}{c}{Kolmogorov-Smirnov} && \multicolumn{4}{c}{Cramér-von Mises} &       & \multicolumn{4}{c}{Covariance distance} \\
		\cmidrule{3-6}\cmidrule{8-11}\cmidrule{13-16}    \multicolumn{1}{c}{p} & n: & $ 100 $   & $ 250 $   & $ 500 $   & $ 1000 $  && $ 100 $   & $ 250 $   & $ 500 $   & $ 1000 $  && $ 100 $   & $ 250 $   & $ 500 $   & $ 1000 $ \\
		\midrule
		1 && $1.000$ & $1.000$ & $1.000$ & $1.000$ && $0.999$ & $1.000$ & $1.000$ & $1.000$ && $0.942$ & $1.000$ & $1.000$ & $1.000$ \\
		2 && $0.743$ & $0.999$ & $1.000$ & $1.000$ && $0.811$ & $1.000$ & $1.000$ & $1.000$ && $0.926$ & $1.000$ & $1.000$ & $1.000$ \\
		3 && $0.204$ & $0.447$ & $0.702$ & $0.949$ && $0.194$ & $0.499$ & $0.804$ & $0.986$ && $0.921$ & $1.000$ & $1.000$ & $1.000$ \\
		4 && $0.131$ & $0.247$ & $0.458$ & $0.722$ && $0.127$ & $0.280$ & $0.514$ & $0.823$ && $0.917$ & $1.000$ & $1.000$ & $1.000$ \\
		5 && $0.105$ & $0.165$ & $0.272$ & $0.472$ && $0.100$ & $0.174$ & $0.324$ & $0.571$ && $0.911$ & $1.000$ & $1.000$ & $1.000$ \\
		\bottomrule
	\end{tabular}%
	}
	\caption{\label{tab:curse_power} Empirical power under the alternative hypothesis, with $\alpha = 0.05$, testing the CAR($p$) process with order $p \in \lbrace 1,2,3,4,5 \rbrace$.}%
\end{table}%


\section{Real data application} \label{sec:5}

\begin{figure}[t]
	\centering
	\includegraphics[width=\linewidth]{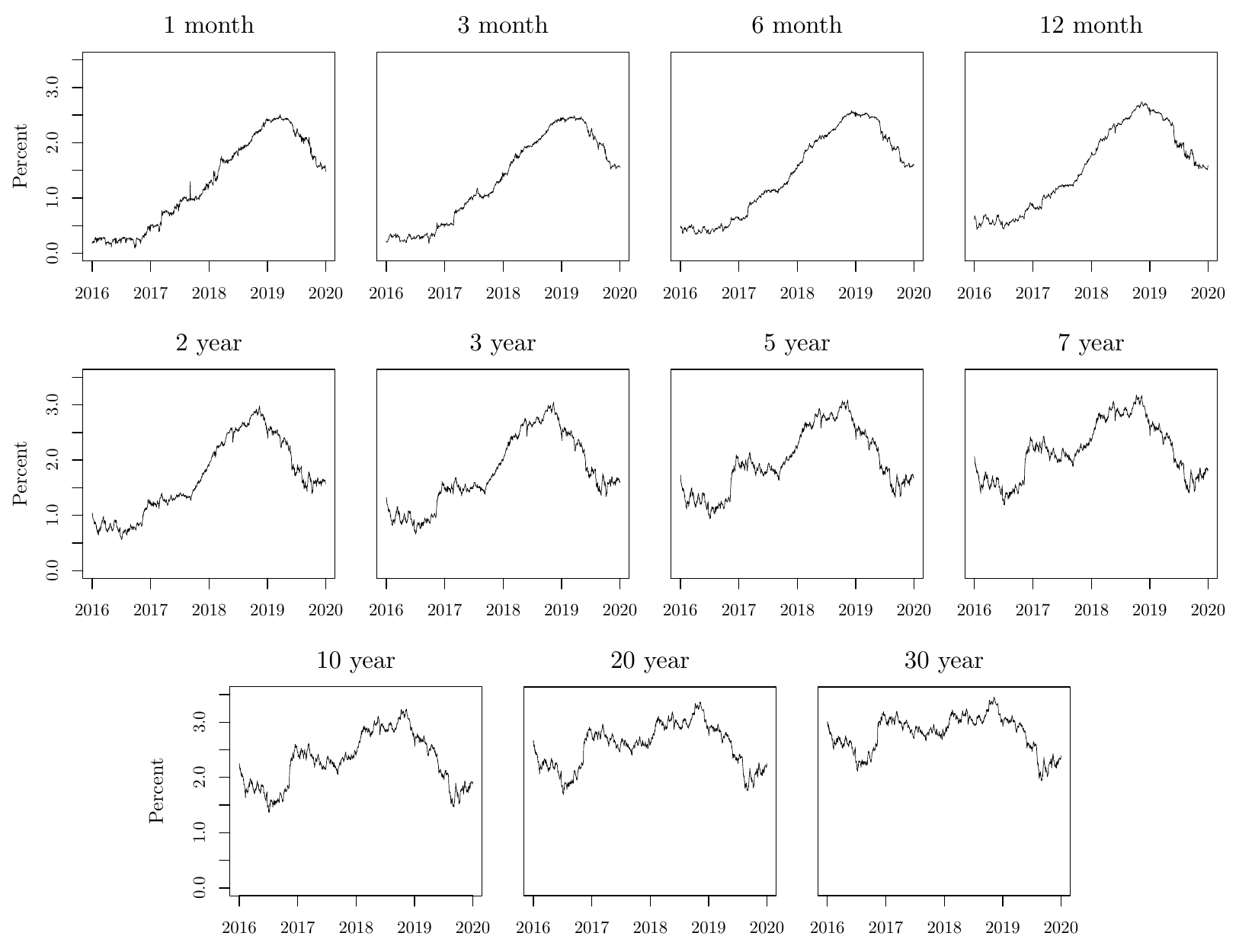}
	\caption{U.S. daily Treasury Securities rate with different maturities.}
	\label{fig:Tbill}
\end{figure}

We apply the methodology reviewed in Section~\ref{sec:2} and Section~\ref{sec:3} to interest rate models involving Treasury securities from the secondary market rates. We use daily market yields on U.S. Treasury securities with different maturities, ranging from 1 month to 30 years, obtained from the Federal Reserve Bank of St. Louis. Treasury securities (T-bills, T-notes, and T-bonds) are financial contracts issued by the U.S. government with price $P(t, \tau)$ at time $t$ that yields a know amount on the maturity date $\tau$, where $P(t,\tau)$ is determined by the rate evolution. The data consist of 999 daily observations ($\Delta = 1/252$), from January 2, 2016 to December 31, 2019, for 11 maturities (see Figure~\ref{fig:Tbill}).

%

As in the previous section, we consider the empirical regression (ER) process-based test of \cite{monsalve2011goodness} and its asymptotically distribution-free (ADF) transformation \citep{chen2015asymptotically}, for both Kolmogorov-Smirnov (KS) and Cramér-von Mises (CvM) test statistics; the nonparametric (NP) proposal of \cite{gao2008specification}; and the test based on distance covariance (DCOV) introduced in Section~\ref{sec:3}. We apply the procedures to test the parametric specification of the diffusion function of CKLS interest rate model \citep{chan1992empirical} given by
\begin{equation*}
\dif X_t = \kappa (\mu - X_t) + \sigma X_t^\gamma \dif W_t,
\end{equation*}
that is, the null hypothesis $\mathcal{H}_0 \colon \sigma (X_t, \boldsymbol{\theta}) = \sigma X_t^\gamma$. Table~\ref{tab:pvalues} reports the $p$-values for the goodness-of-fit tests, which were obtained based on $1\;000$ bootstrap resamples. The empirical $p$-values are large for longer maturities (7, 10, 20 and 30 years), hence, the samples show no significant evidences against the CKLS diffusion function for any sensible significance level.
Conversely, the $p$-values lead to strong rejections of the null hypothesis for short maturities (1,3,6 and 12 months), implying that the model is inadequate to explain the volatility of the T-bill series. The 2, 3 and 5 years maturities show the most noticeable disparity among the different procedures. The deterministic parametric specification of the CKLS diffusion function, where many known models can be nested (as \citealp{vasicek1977equilibrium}, with $\gamma=0$ or the \citealp{cox1985intertemporal} model with $\gamma=1/2$), is not able to explain the dynamics of the treasury data for short maturities. Therefore, capturing the volatility behavior may require a more intricate form, such as stochastic volatility.

\begin{table}[H]
	\centering
	\begin{tabular}{lcccccc}
		\toprule
		&   \multicolumn{2}{c}{ER} & \multicolumn{2}{c}{ADF} &  &  \\
		\cmidrule(l){2-3}\cmidrule(l){4-5} 
		\multicolumn{1}{c}{Maturity} & KS    & CvM   & KS & CvM  & NP & DCOV \\
		\midrule
		\textit{1 month} & $< 0.001$ & $< 0.001$ & $< 10^{-6} $ & $< 10^{-6} $ & $< 0.001$ & $0.006$ \\
		\textit{3 month} & $0.002  $ & $0.014  $ & $5.9 \times 10^{-4}$ & $0.002  $ & $0.013  $ & $0.003$ \\
		\textit{6 month} & $0.003  $ & $0.017  $ & $0.008$ & $0.038  $ & $0.001  $ & $0.002$ \\
		\textit{12 month}& $< 0.001$ & $< 0.001$ & $< 10^{-6} $ & $< 10^{-6} $ & $< 0.001$ & $0.002$ \\
		\textit{2 year}  & $0.007  $ & $0.013  $ & $2 \times 10^{-4}$ & $3.2 \times 10^{-4}$ & $0.001  $ & $0.056$ \\
		\textit{3 year}  & $0.456  $ & $0.662  $ & $0.004  $ & $0.008  $ & $0.875  $ & $0.164$ \\
		\textit{5 year}  & $0.557  $ & $0.680  $ & $0.078  $ & $0.170  $ & $0.052  $ & $0.428$ \\
		\textit{7 year}  & $0.847  $ & $0.926  $ & $0.447  $ & $0.590  $ & $0.280  $ & $0.532$ \\
		\textit{10 year} & $0.906  $ & $0.967  $ & $0.969  $ & $0.922  $ & $0.987  $ & $0.469$ \\
		\textit{20 year} & $0.981  $ & $0.983  $ & $0.380  $ & $0.400  $ & $0.998  $ & $0.436$ \\
		\textit{30 year} & $0.549  $ & $0.847  $ & $0.614  $ & $0.805  $ & $0.999  $ & $0.488$ \\
		\bottomrule
	\end{tabular}%
	\caption{\label{tab:pvalues} $p$-values for the goodness-of-fit test for the CKLS parametric form of the diffusion function.}%
\end{table}%

\begin{figure}[t]
	\centering
	\includegraphics[width=0.9\linewidth]{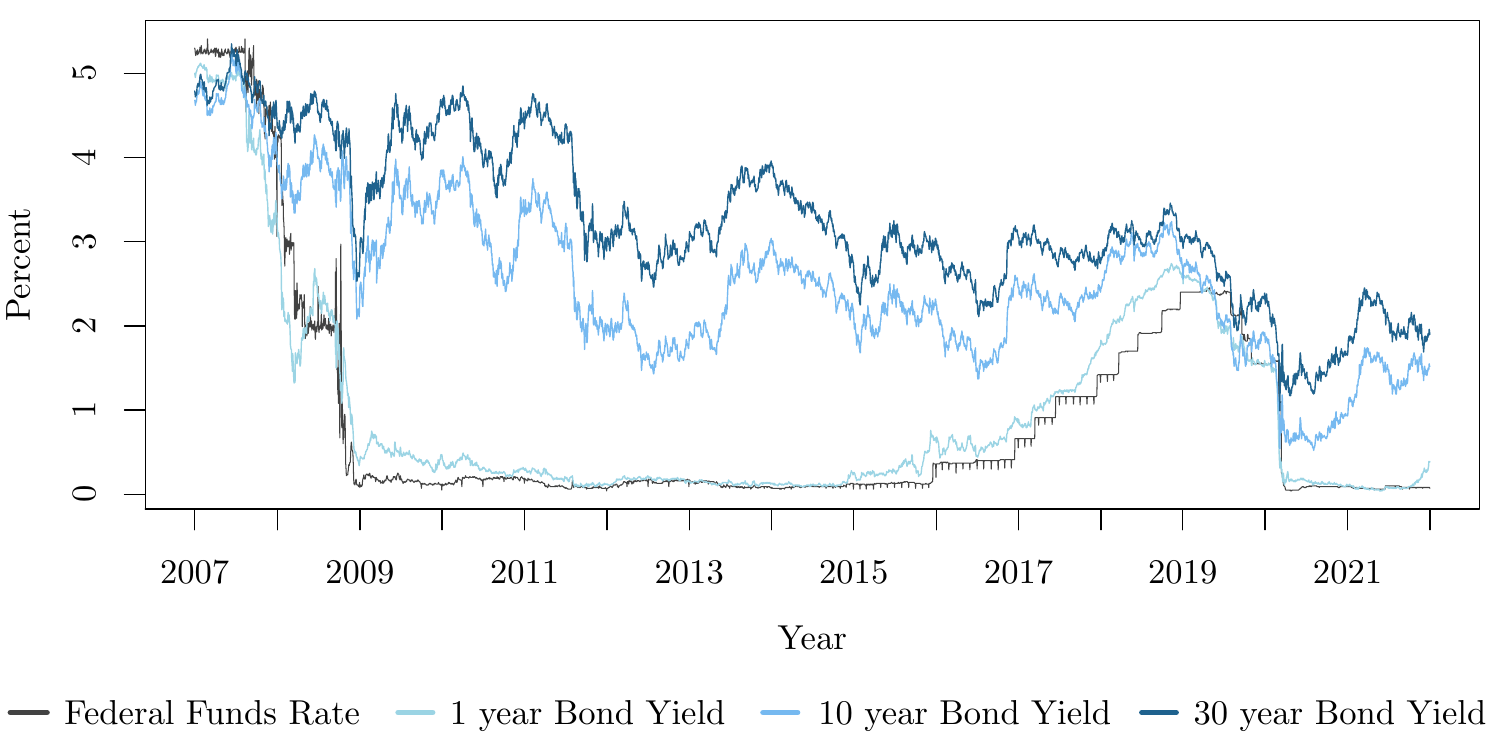}
	\caption{U.S. federal funds rate and U.S. treasury yields (1, 10 and 30 year bond yields).}
	\label{fig:Tbill_rate}
\end{figure}

Under the sample period considered, central banks from different economies have acted in unprecedented scale. Regarding the U.S., the Federal Reserve has increased rates several times since 2015 and reduced the rate three times in 2019, for a total of 75 basis points. Short maturities of treasury yield are tied close to Fed policy, as shown in Figure~\ref{fig:Tbill_rate}, where the evolution of the one year bond yield moves closely with the Federal fund rate. Therefore, the CKLS model might fail to adequately account structural shifts of the process due to intervention of the Federal Reserve and extending the diffusion function to incorporate unexpected shocks could provide a superior empirical fit.




\begin{figure}[t!]
	\centering
	\includegraphics[width=\linewidth]{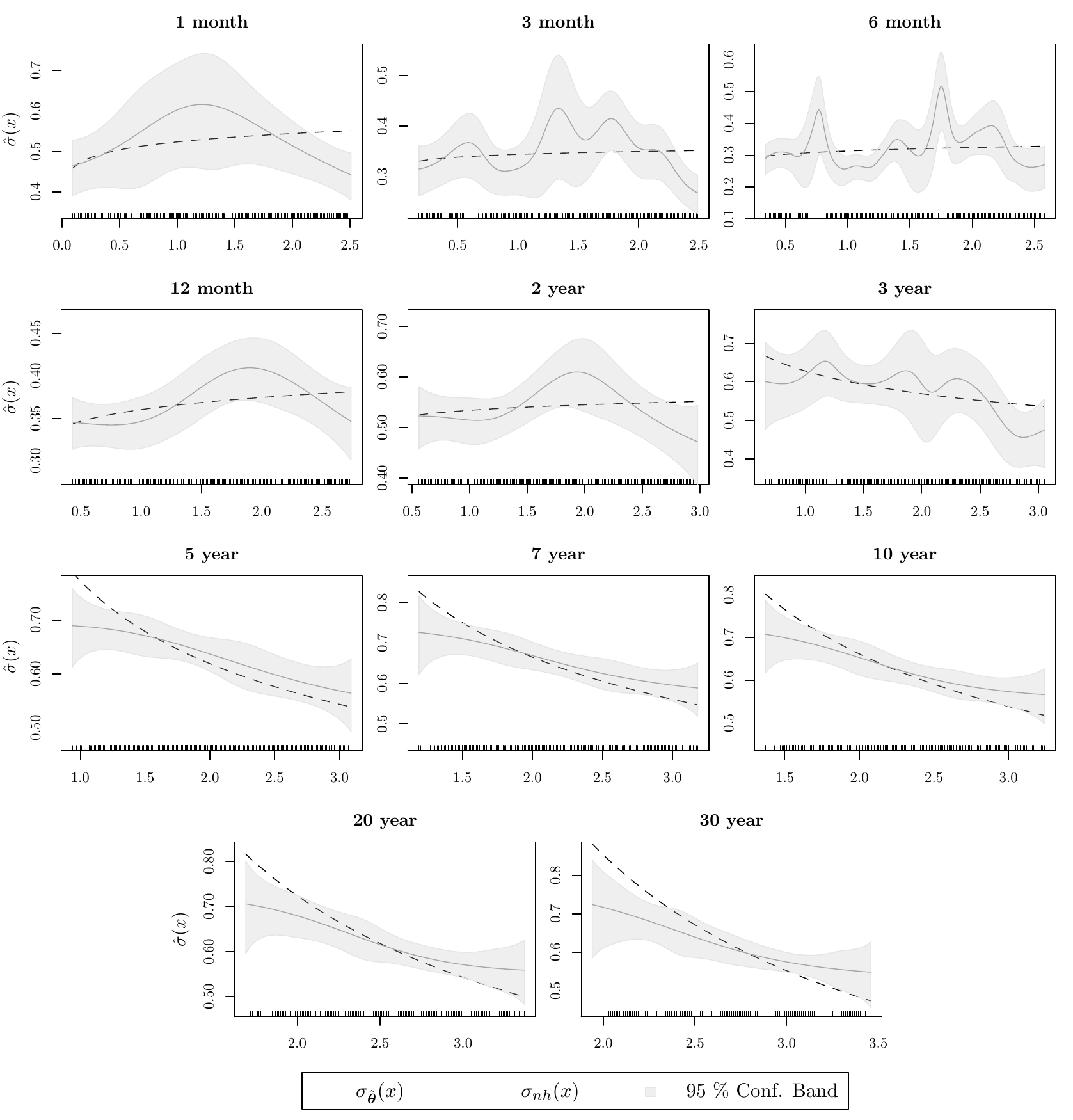}
	\caption{Parametric (dashed line) and nonparametric (gray solid line) estimation of the volatility function for the U.S. daily Treasury Securities. Shaded area indicates a 95\% pointwise confidence interval for the diffusion function estimation.}
	\label{fig:Tbill_est}
\end{figure}

To provide more insight into the rejection of the null hypothesis for shorter maturities (T-bill), we compare the parametric and non parametric estimation of the volatility function for all maturities. Given the kernel estimate $\sigma_{nh}^2(x)$ in~\eqref{eq:sigma_nh} of $\sigma^2(\cdot)$, we have \citep{hardle1992applied}
\begin{equation*} 
\sqrt{nh} \Big( \sigma^2_{nh} (x) - \mathbb{E} \left[ \sigma^2_{nh} (x) \mathrel{\big|} X_{t_0}, \dots, X_{t_n} \right] \Big) \xrightarrow{\;\, d \;\,} N \bigg( 0, \frac{R(K) V(u_{t_i}^2 \mathrel{|} X_{t_i} = x)}{\pi(x)} \bigg),
\end{equation*}
with $R(K) = \int K^2 (s) \dif s$, $u^2_{t_i} = \left(X_{t_{i+1}} - X_{t_i} \right)^2 / \Delta - \sigma^2(X_{t_i})$ and $\pi(x)$ the marginal density function. The asymptotic variance is estimated by plugging the kernel density estimator $\pi_{nh}(x)$ in~\eqref{eq:pi_nh} as an estimate of $\pi(x)$ and estimating
%
%
\begin{equation*}
\hat{V} \left(u^2_{t_i} \mid X_{t_i} = x \right) = \sum_{j=1}^{n-1} W_{nj}(x) \Delta^{-1} \left(X_{t_{i+1}} - X_{t_i} \right)^4 - \left(\sigma_{nh}^2 (x) \right)^2.
\end{equation*}
Thus, $(1-\alpha)$ pointwise confidence bands for the diffusion function $\sigma(\cdot)$ are given by $\big[ \big( \sigma_{nh}^2 (x) \pm z_{1-\alpha/2} \sqrt{\hat{V}_{\sigma^2} (x)/nh} \big)^{1/2} \big]$, with $z_{\alpha}$ the $\alpha$ quantile of the standard Gaussian distribution and $\hat{V}_{\sigma^2} (x) = R(K) \hat{V}(u_{t_i}^2 \mathrel{|} X_{t_i} = x)/\pi_{nh}(x)$. 
The parametric and kernel estimation are shown in Figure~\ref{fig:Tbill_est}, along 95\% confidence bands, where a Gaussian kernel was used, $K(u) = (2\pi)^{-1/2} \exp (-u^2/2) $, with a leave-one-out cross-validation bandwidth.


The poor fit of the parametric diffusion function for T-bills, almost constant, may be due to the noisy nature of the series or the presence of jumps and structural shocks. In the parametric specification of the CKLS model, the parameter $\gamma$ is called the ``level effect'', as it determines the sensitivity of the variance with respect to the interest rate level. The estimated value of $\gamma$ is close to zero for the shorter yields and decreases with maturity, reaching negative values. In this regard, the models may be unstationary, as $\gamma < 0$ does not guarantee stationarity. The kernel estimation for longer maturities, very close to the parametric estimation, also shows an inverse relationship between the interest rate level and the volatility, which is rather unexpected. However, the null hypothesis was not rejected. As already discussed, the sample period considered was subject of structural changes in markets. The interest rates on some long maturities bonds fell bellow short-term debt, as shown in Figure~\ref{fig:Tbill_rate}, where the spread between the 1 year Treasury note and the 10 year bond inverted in 2019. Long-term bond yields evolved in the direction of shorter terms, with the 30 year bond falling bellow 2\%, reaching its lowest level up to that date. This implied a flattening yield curve for the sample period, actually inverting in 2019. Higher yields were pushed down by several factors other than market expectations about interest rate evolution, so that the inversion of the yield curve may be a result of structural market changes, which could account for the negative level effect in long-term bonds observed in the sample.

\section{Conclusions} \label{sec:6}

We reviewed goodness-of-fit test for diffusion models, collecting the proposals of the last 25 years, many of which are extension of more classical proposals to the context of continuous-time stochastic processes. An application of correlation distance ideas to develop a test for the parametric specification of diffusion models was also provided. Empirical results showed that the performance of the procedures, in terms of size and power, are close, calibrating adequately the null hypothesis and detecting a variety of alternatives. However, when dimension increases, some tests lose power. In this scenario, the correlation distance proposal outperforms others, avoiding the curse of dimensionality. An application of the tests to interest rate models involving U.S. daily Treasury securities was also undertaken, finding empirical support to the CKLS diffusion function for long maturities. As the maturity range was reduced, the $p$-values became small, indicating  that the model was mis-specified for T-bill (1,3,6 and 12 month maturities) series.




%

\bibliography{biblio}
\bibliographystyle{apalike} 

\end{document}